\documentclass[pre,preprint,amsmath,amssymb]{revtex4}

\usepackage{graphicx}

\newcommand{\lang}{\left\langle}
\newcommand{\rang}{\right\rangle}

\begin{document}

\title{Path integral analysis of Jarzynski's equality: Analytical results}

\author{David D. L. Minh}
\email{daveminh@gmail.com}
\author{Artur B. Adib}
\email{adiba@mail.nih.gov}
\affiliation{
Laboratory of Chemical Physics, NIDDK, National Institutes of Health, Bethesda, Maryland 20892-0520, USA
}

\date{\today}

\begin{abstract}
We apply path integrals to study nonequilibrium work theorems in the context of Brownian dynamics, deriving in particular the equations of motion governing the most typical and most dominant trajectories. For the analytically soluble cases of a moving harmonic potential and a harmonic oscillator with time-dependent natural frequency, we find such trajectories, evaluate the work-weighted propagators, and validate Jarzynski's equality.
\end{abstract}

\maketitle


As nonequilibrium work theorems describe exact relationships between path averages and equilibrium thermodynamic properties \cite{jarzynski97-prl,crooks99}, path integrals offer a natural framework for their analysis. Previous studies of such theorems have invoked path integrals primarily with the intent of proving their validity, e.g. by establishing an exact relation between the path integration measures of processes taking place in opposite directions \cite{crooks-thesis,chernyak06}, or by invoking the Feynman-Kac formula for the work-weighted density \cite{hummer01-pnas,hummer05}.  In either case, the theorems can be proved without explicitly computing some of the quantities that make path integrals most useful, for example most likely trajectories and propagators.

Here we complement these efforts by using path integrals to obtain trajectories and work-weighted propagators, and explicitly evaluating these objects for analytically tractable models.  We derive the equations of motion governing two important trajectories (Eq.~(\ref{eq:E-L})): the most typical, which has the highest probability; and the most dominant, which contributes most significantly to the Jarzynski average.  Interest in these trajectories stems in part from the observation that when typical and dominant trajectories diverge, the Jarzynski free energy estimate converges slowly \cite{jarzynski06-pre}.  Path integrals also offer a convenient framework for the computation of work-weighted propagators.  We show that these propagators are analytically soluble for harmonic oscillators with moving equilibrium centers or changing frequencies.  Collectively, our results demonstrate that the utility of path integrals in the context of nonequilibrium work theorems goes beyond simply proving their validity.

We begin our analysis by formulating a path integral representation of the propagator, which for simplicity we take to be that of a one-dimensional overdamped Brownian particle moving on a time-dependent potential $U(x,t)$. We assume a constant diffusion coefficient $D$ and measure energy in units of $k_B T$. Accordingly, the propagator $p = p(x,t|x_0,0)$ from the point $x_0$ at $t=0$ to the point $x$ at time $t$ satisfies the Smoluchowski equation \cite{zwanzig01},
\begin{equation} \label{eq:p}
  \frac{\partial p}{\partial t} = D \frac{\partial}{\partial x} \left[ e^{-U(x,t)} \frac{\partial}{\partial x} \left( e^{U(x,t)} p \right) \right].
\end{equation}
As with the case where $U$ is time-independent \cite{zwanzig01}, we can apply the substitution $p = e^{-U(x,t)/2} q$, where $q = q(x,t|x_0,0)$, to write the equation in a Schr\"odinger-like form,
\begin{equation} \label{eq:q-master}
  \frac{\partial q}{\partial t} = D \frac{\partial^2 q}{\partial x^2} + \left[ \mathcal{V}(x,t) +\frac{1}{2} \frac{\partial U}{\partial t} \right] \, q.
\end{equation}
The ``effective potential'' $\mathcal{V}$ is given by
\begin{equation} \label{eq:V}
  \mathcal{V}(x,t) = D \left[ \frac{U''(x,t)}{2} - \left( \frac{U'(x,t)}{2} \right)^2  \right],
\end{equation}
with primes denoting partial derivatives with respect to $x$.  Expressing the propagator in the form of Eq.~(\ref{eq:q-master}) allows us to directly apply the Feynman-Kac formula \cite{chaichian-book01} to write its solution as a path integral, 
\begin{equation} \label{eq:q}
  q = \int_{x_0,0}^{x,\tau} \! \! dx(t) \, e^{-\int_0^\tau \! dt \left[ \frac{\dot{x}^2}{4D} - \mathcal{V} - \frac{1}{2} \frac{\partial U}{\partial t} \right]},
\end{equation}
where the density $q=q(x,\tau|x_0,0)$ is at time $t=\tau$, $\mathcal{V} \equiv \mathcal{V}(x(t),t)$, and analogously for $U$. (In this notation, the Wiener integral over all trajectories that start at $x_0$ at $t=0$ and end at $x$ at $t=\tau$ is obtained by setting $\mathcal{V}=U=0$ in the above equation, cf. \cite{chaichian-book01}). Returning to the original variable $p = p(x,\tau|x_0,0)$ and imposing the boundary condition $p(x,0|x_0,0) = \delta(x-x_0)$, we arrive at the desired path integral representation of the propagator, 
\begin{equation} \label{eq:p-pathint}
  p = e^{-\frac{\Delta U}{2}} \int_{x_0,0}^{x,\tau} \! \! dx(t) \, e^{-\int_0^\tau \! dt \left[ \frac{\dot{x}^2}{4D} - \mathcal{V} - \frac{1}{2} \frac{\partial U}{\partial t} \right]},
\end{equation}
where $\Delta U \equiv U(x,\tau) - U(x_0,0)$. Note that our approach differs from previous treatments \cite{crooks-thesis,chernyak06} in that no stochastic integrals (e.g. Stratonovic or Ito) are used; through the Feynman-Kac formalism, only ordinary time integrals and the usual Wiener contribution appear in the path integral. This derivation thus avoids a common confusion associated with the definition of the action due to such stochastic integrals \cite{adib08-jpcb}.

To make contact with Jarzynski's equality, consider the {\em work-weighted propagator} $p_w = p_w(x,\tau|x_0,0)$, defined as
\begin{align} 
  p_w & \equiv e^{-\frac{\Delta U}{2}} \int_{x_0,0}^{x,\tau} \! \! dx(t) \, e^{-\int_0^\tau \! dt \left[ \frac{\dot{x}^2}{4D} - \mathcal{V} - \frac{1}{2} \frac{\partial U}{\partial t} \right]} \, e^{-w[x(t)]} \nonumber \\
      & = e^{-\frac{\Delta U}{2}} \int_{x_0,0}^{x,\tau} \! \! dx(t) \, e^{-\int_0^\tau \! dt \left[ \frac{\dot{x}^2}{4D} - \mathcal{V} + \frac{1}{2} \frac{\partial U}{\partial t} \right]}. \label{eq:pw}
\end{align}
This propagator differs from Eq.~(\ref{eq:p-pathint}) in that each trajectory in the path integral is endowed with the additional weight $e^{-w[x(t)]}$, where $w[x(t)] = \int_0^\tau \! dt \, ( \partial U / \partial t)$ is the work along the trajectory $x(t)$ \cite{jarzynski97-prl}.  The path average in Jarzynski's equality, $\langle e^{-w} \rangle = Z_\tau / Z_0$, can be written in terms of $p_w$ as
\begin{equation} \label{eq:jarz-avg}
  \lang e^{-w} \rang = \int dx_0 \int dx \, \frac{e^{-U(x_0,0)}}{Z_0}  p_w(x,\tau|x_0,0),
\end{equation}
where $Z_t \equiv \int dy \, e^{-U(y,t)}$. Thus, knowledge of the analytical form of $p_w$ reduces the computation of the Jarzynski average $\langle e^{-w} \rangle$ to simple quadrature.

A unique feature of path integrals is that they allow one to extract noteworthy trajectories from a given path average, such as Eq.~(\ref{eq:jarz-avg}). This is done by first identifying the action that dictates the trajectory probability and then optimizing the desired functional (see \cite{orland06,orland07} for a similar approach in the context of time-independent potentials).  In the case of Eq.~(\ref{eq:jarz-avg}), the path average can be written as 
\begin{equation}
  \lang e^{-w} \rang = \int \! dx_0 \int \! dx \int_{x_0,0}^{x,\tau} \! \! dx(t) \, \frac{e^{-S[x(t)]}}{Z_0} \, e^{-w[x(t)]},
\end{equation}
where the action functional is defined as
\begin{equation}
  S[x(t)] \equiv { \textstyle \frac{U(x,\tau) + U(x_0,0)}{2} + \int_0^\tau \! dt \left[ \frac{\dot{x}^2}{4D} - \mathcal{V} - \frac{1}{2} \frac{\partial U}{\partial t} \right] }.
\end{equation}
Notably, in contrast to the usual case where the end-points $x_0$ and $x$ are fixed, here the action embodies the probability of observing the end-points themselves; upon minimization, this gives rise to the ``natural'' boundary conditions below. Two trajectories of interest are the {\em most typical} ($x_T(t)$) and the {\em most dominant} ($x_D(t)$).  In the present formalism, they correspond to the trajectories that minimize the action $S$ alone and that minimize the combined action-work functional, $S+w$, respectively. Carrying out the functional optimization of $S$ and $S+w$ with free end-points, we obtain the Euler-Lagrange equations
\begin{equation} \label{eq:E-L}
  \frac{\ddot{x}(t)}{2D} = - \frac{\partial}{\partial x} \left[ \mathcal{V} \pm \frac{1}{2} \frac{\partial U}{\partial t} \right],
\end{equation}
subject to the natural boundary conditions $\dot{x}(0)  = D \left( \partial U / \partial x \right)_{t=0}$ and $\dot{x}(\tau) = -D \left( \partial U / \partial x \right)_{t=\tau}$. The sign of $\partial U/\partial t$ in Eq.~(\ref{eq:E-L}) is positive for the most typical and negative for the most dominant trajectories, whereas the boundary conditions are the same for both types.

The path integral formulation of the work-weighted propagator (Eq.~(\ref{eq:pw})) and the Euler-Lagrange equations for the most typical and most dominant trajectories (Eq.~(\ref{eq:E-L})) are the central objects of interest in this paper.  We now demonstrate them on two analytically tractable problems.

\begin{figure}
\begin{center}
\includegraphics[width=250pt]{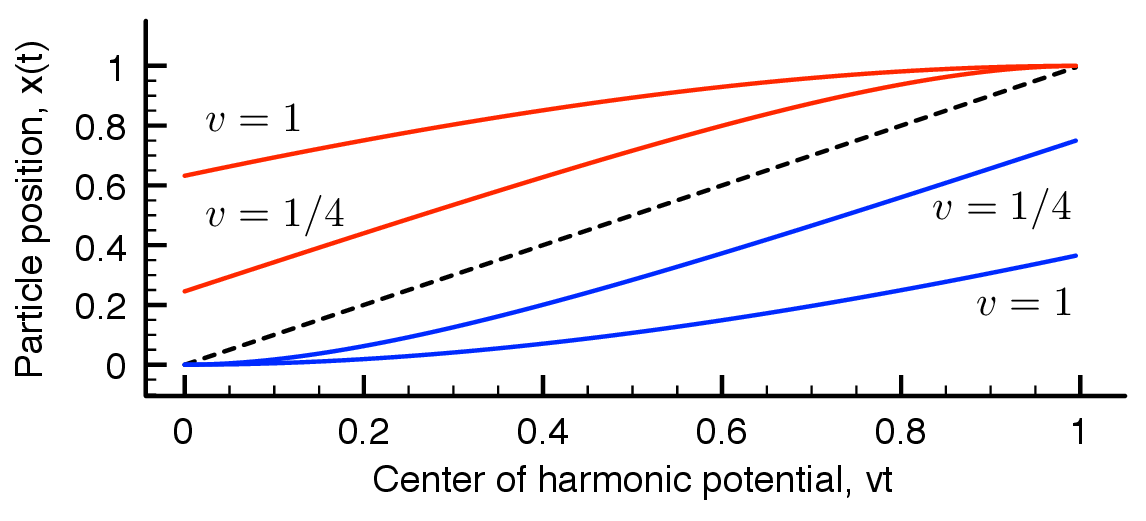} 
\caption{Illustration of Eqs.~(\ref{xT})-(\ref{xD}). The most typical (blue, bottom two curves) and most dominant (red, top two) trajectories for two different velocities are plotted with respect to the center of the harmonic well, $vt$ (dashed).  The parameters are $v\tau = D = k = 1$.}
\label{fig:trajs}
\end{center}
\end{figure}

%
%

{\em (a) Moving oscillator}, $U(x,t) = k(x-vt)^2/2$. Here the spring constant $k \geq 0$ is time-independent, and the center of the harmonic potential moves at constant velocity $v$. We first compute the most typical and most dominant trajectories. The Euler-Lagrange equations above become
\begin{equation}
  \ddot{x} = (Dk)^2 (x-vt) \pm Dk v,
\end{equation}
with boundary conditions $\dot{x}(0) = Dk\, x(0)$ and $\dot{x}(\tau) = -Dk(x(\tau) - vt)$.  The sign convention is the same as Eq.~(\ref{eq:E-L}). Solving these elementary differential equations, we find
\begin{align}
  x_T(t) & = vt - { \textstyle \frac{v}{Dk} } \left( 1 - e^{-Dkt} \right), \label{xT}  \\
  x_D(t) & = vt + {\textstyle \frac{v}{Dk} } \left( 1 - e^{-Dk(\tau - t)} \right). \label{xD}
\end{align}
Thus, in the allotted time interval $0\leq t \leq \tau$, the most typical trajectory starts out aligned with the center of the well and lags behind at later times, while the most dominant one starts ahead of the center in the beginning, but falls back at the final time $\tau$ (see Fig.~\ref{fig:trajs}).

We now proceed to the computation of the work-weighted propagator. For the moving harmonic potential, Eq.~(\ref{eq:pw}) gives
\begin{equation}
  p_w = e^{-\frac{\Delta U}{2} + \frac{Dk\tau}{2}} \\
  \times \int_{x_0,0}^{x,\tau} \! \! dx(t) \, e^{-\int_0^\tau dt \left[ \frac{\dot{x}^2}{4D} + \frac{Dk^2}{4}(x-vt)^2 - \frac{vk}{2}(x-vt) \right]}
\end{equation}
This path integral can be simplified with the change of variables $z(t) = x(t) - vt - \frac{v}{Dk}$, which does not change the integration measure. This gives
\begin{equation} \label{eq:pw-moving1}
  p_w = e^{-\frac{\Delta U}{2} + \frac{Dk\tau}{2} -\frac{v}{2D}(x - x_0 - v\tau)} \, G(z,\tau|z_0,0),
\end{equation}
where $z_0 = x_0 - \frac{v}{Dk}$, $z = x - v\tau - \frac{v}{Dk}$, and $G$ is the Gaussian path integral
\begin{equation}
  G(z,\tau|z_0,0) = \int_{z_0,0}^{z,\tau} \! \! dz(t) \, e^{-\int_0^\tau dt \left[ \frac{\dot{z}^2}{4D} + \frac{Dk^2}{4}z^2 \right]}.
\end{equation}
This is the propagator of a Brownian particle subject to a harmonic, time-independent source/sink term. Its result is well-known (cf. \cite{chaichian-book01}), namely
\begin{equation}
  G(z,\tau|z_0,0) = \frac{ \exp\left[ -\frac{k}{4}\frac{(z^2 + z_0^2)\cosh(Dk\tau) - 2 z z_0}{\sinh (Dk\tau)} \right] } { \sqrt{\frac{4\pi}{k} \sinh(Dk\tau)} }.
\end{equation}
Finally, using this result in Eq.~(\ref{eq:pw-moving1}) and expressing the final formula in terms of the convenient variables $z$ and $z_0$ (see above), we obtain
\begin{equation} \label{eq:pw-moving-final}
  p_w = \frac{ \exp\left[ -\frac{k}{2}\frac{(z-z_0e^{-Dk\tau})^2}{1-e^{-2Dk\tau}} -\frac{v}{D}(z-z_0)  \right]  } { \sqrt{\frac{2\pi}{k}(1-e^{-2Dk\tau})} }.
\end{equation}
In accordance with expectations, when $v=0$, this expression reduces to the well-known formula for the propagator of a Brownian particle in a stationary harmonic potential (Ornstein-Uhlenbeck process, see e.g. \cite{risken}), as in this case the work is identically zero and we must have $p=p_w$.  Additionally, using this propagator in Eq.~(\ref{eq:jarz-avg}) and performing the two Gaussian integrals leads to $\langle e^{-w} \rangle = 1$, which confirms the Jarzynski equality prediction.

We leave this section with a comparison of our results with those of Mazonka and Jarzynski \cite{mazonka99}. These authors directly solved a Fokker-Planck equation for the joint work-position density $f(y,w|y_0)$, where in their study $y(t) = x(t) - vt$. Integrating over the work, the marginal density of $y(t)$ was found to be Gaussian with a mean moving according to a simple expression (Eq.~(14a) of that reference).  Incidentally, upon Boltzmann-averaging over the initial condition $y_0$, this reduces to our Eq.~(\ref{xT}).  We note, however, that this coincidence between average position and most typical trajectory is not to be expected in general, and is likely to be a peculiarity of harmonic problems.  Lastly, we note that Eq.~(\ref{eq:pw-moving-final}) can in principle be obtained from the full expression for $f(y,w|y_0)$ in the appendix of Ref.~\cite{mazonka99} by integrating out $w$ from $f$ with weight $e^{-w}$.

%
%

{\em (b) Time-dependent spring constant}, $U(x,t) = k(t) x^2/2$, where $k(t)\geq0$. This problem has been used in the numerical examples of Ref.~\cite{jarzynski97-pre}. Perhaps unsurprisingly, the most typical and most dominant trajectories are found to be $x_T(t) = x_D(t) = 0$ for any $k(t)$. This simply reflects the parity symmetry of the problem, and indicates that other properties, such as the width of the instantaneous position distribution, are responsible for the observed convergence properties of the Jarzynski average \cite{jarzynski97-pre}.

A less trivial aspect of this problem is its work-weighted propagator. From Eq.~(\ref{eq:pw}), we obtain
\begin{equation} \label{pw-spring}
  p_w = e^{-\frac{\Delta U}{2} + \frac{D}{2} \int_0^\tau \! dt \, k} \\
  \times \int_{x_0,0}^{x,\tau} \! \! dx(t) \, e^{-\int_0^\tau \! dt \, \left[ \frac{\dot{x}^2}{4D} + \frac{\dot{k} +Dk^2}{4} x^2 \right]},
\end{equation}
where $k = k(t)$. Due to the fortuitous combination $\dot{k} +Dk^2$ in the effective harmonic potential, this propagator can be computed in closed analytical form for arbitrary $k(t)$. To proceed, we factor out the $x_0,x$ dependence from the path integral with the usual substitution for harmonic potentials \cite{feynman98}, $x(t) = \overline{x}(t) + y(t)$.  The trajectory $\overline{x}(t)$ is an extremum of the action in Eq.~(\ref{pw-spring}), satisfying the Euler-Lagrange equation
\begin{equation} \label{xbar}
  \ddot{\overline{x}} = (D\dot{k} + D^2 k^2 ) \, \overline{x},
\end{equation}
with boundary conditions $\overline{x}(0) = x_0$ and $\overline{x}(\tau) = x$, i.e. $y(0) = y(\tau) = 0$. After integration by parts and changing the path integration variable from $x(t)$ to $y(t)$, we get
\begin{equation} \label{eq:pw-G}
  p_w = e^{ -\frac{\Delta U}{2} + \frac{D}{2} \int_0^\tau \! dt \, k } \, e^{- \left. \frac{ \overline{x} \, \dot{\overline{x}} } {4D} \right|_0^\tau } \, G(\tau),
\end{equation}
where we have defined the $x_0,x$-independent quantity
\begin{equation}
  G(\tau) = \int_{0,0}^{0,\tau} \! \! dy(t) \, e^{-\int_0^\tau \! dt \left[ \frac{\dot{y}^2}{4D} + \frac{\dot{k} +Dk^2}{4} y^2 \right] }.
\end{equation}
This Gaussian path integral can be formally evaluated in different ways. Using the Gelfand-Yaglom method \cite{chaichian-book01}, we find $G(\tau) = 1/\sqrt{4 \pi D \varphi(\tau)}$, where $\varphi(t)$ satisfies the same differential equation as $\overline{x}(t)$ (Eq.~(\ref{xbar})), with boundary conditions $\varphi(0) = 0$ and $\dot{\varphi}(0) = 1$.

Thus, we have reduced the evaluation of the path integral in Eq.~(\ref{pw-spring}) to the solution of a single differential equation (Eq.~(\ref{xbar})). There are two independent solutions to this equation, namely
\begin{align}
  \psi_1(t) & = e^{D \int_0^t ds k}, \label{psi1} \\
  \psi_2(t) & = e^{D \int_0^t ds k} \, \left[ {\textstyle \int_0^t \! du \, e^{-2D \int_0^u \! ds \, k} } \right] . \label{psi2}
\end{align}
The desired quantities $\overline{x}(t)$ and $\varphi(t)$ are obtained by a linear combination of $\psi_1$ and $\psi_2$ with the appropriate boundary conditions. Here we omit the details of this computation and simply quote the final result for $p_w$, obtained by plugging these solutions into Eq.~(\ref{eq:pw-G}):
\begin{equation} \label{pw-koft}
  p_w = 
    \textstyle
    {
      \sqrt{ \frac{\psi_1}{4 \pi D \psi_2} }
      \exp \left\{ -\frac{1}{4D} \left[ \left( 2Dk(\tau) + \frac{1}{\psi_1 \psi_2} \right)x^2 \right. \right.
    } \\
    \textstyle
    {
      \left. \left. -\left(2Dk(0) - \frac{\psi_1}{\psi_2}\right)x_0^2 - \frac{2}{\psi_2} x x_0 \right] \right\},
    }
\end{equation}
where for brevity of notation we have defined $\psi_1 \equiv \psi_1(\tau)$ and $\psi_2 \equiv \psi_2(\tau)$. Thus, for any $k(t)$, the work-weighted propagator is a Gaussian function whose time-dependent coefficients can be found via quadrature, i.e. via Eqs.~(\ref{psi1})-(\ref{psi2}).
As in the previous problem, one can check that this result reduces to the propagator for a one-dimensional Ornstein-Uhlenbeck process when $k(t)$ is a constant. Furthermore, upon using this expression for $p_w$ in Eq.~(\ref{eq:jarz-avg}) and carrying out the two Gaussian integrals, massive cancelation of terms ensues, leaving $\langle e^{-w} \rangle = \sqrt{k(0)/k(\tau)}$, which also validates the Jarzynski equality prediction.

%
%

Lastly, we consider work-weighted propagators and dominant/typical trajectories in the context of forward and reverse processes, in the sense of Crooks \cite{crooks99}.  For a given time-dependent energy function $U(x,t)$ defined between times $t=0$ and $t=\tau$, let us arbitrarily call the dynamics under the potential $U(x,t)$ the {\em forward} process, and the dynamics under the time-reversed potential $U(x,\tau-t)$ the {\em reverse} process. By definition, the forward propagator $p^F \equiv p^F(x,\tau|x_0,0)$ is given by Eq.~(\ref{eq:p-pathint}), while the reverse propagator $p^R(x,\tau|x_0,0)$ is given by the same expression with the replacement $U(x,t) \to U(x,\tau-t)$.  Now consider the quantity $p^R \equiv p^R(x_0,\tau|x,0)$, i.e. the reverse propagator with inverted initial and final conditions, $x$ and $x_0$ respectively. After the time-reversal change of variables $s = \tau - t$ and $y(s) = x(\tau - s)$ in the path integral, we obtain
\begin{equation} \label{eq:pR}
  p^R = e^{\frac{\Delta U}{2}} \int_{x_0,0}^{x,\tau} \! \! dy(t) \, e^{-\int_0^\tau \! dt \left[ \frac{\dot{x}^2}{4D} - \mathcal{V} + \frac{1}{2} \frac{\partial U}{\partial t} \right]},
\end{equation}
where $\Delta U$ is defined as before. We thus see that $p^R$ differs from $p^F$ simply in the signs of $\Delta U$ and $\partial U/\partial t$.

Two interesting consequences are of immediate notice.  First, the following relation is verified:
\begin{equation} \label{eq:w-detbal}
  e^{-U(x_0,0)} p^F_w(x,\tau|x_0,0) = e^{-U(x,\tau)} p^R(x_0,\tau|x,0),
\end{equation}
where the forward work-weighted propagator $p^F_w$ is given by Eq.~(\ref{eq:pw}).  When $U$ is time-independent, this result reduces to detailed balance, i.e. $e^{-U(x_0)} p(x,\tau|x_0,0) = e^{-U(x)} p(x_0,\tau|x,0)$, as in this case $w=0$ and the process directions are immaterial. This identity can thus be seen as version of detailed balance in the context of time-dependent potentials, and establishes that the computation of $p_w$ is no easier or harder than that of $p$.  In contrast, when $p_w$ is averaged over the initial Boltzmann distribution (by dividing the above equation by $Z_0$ and integrating over $x_0$), the resultant density has the generic closed-form analytical solution $e^{-U(x,\tau)}/Z_0$ \cite{jarzynski97-pre,hummer01-pnas}.  Further integration over $x$ leads to Jarzynski's equality.

Secondly, as per Eq.~(\ref{eq:E-L}) and the definition of reverse process above, the time-reversal of the typical and dominant trajectories in the reverse process satisfy precisely the same Euler-Lagrange equations and boundary conditions of the dominant and typical trajectories in the forward process, respectively. For example, observing Fig.~\ref{fig:trajs} from right to left, the typical trajectories become the dominant ones, and vice-versa. Our results thus provide means to explicitly compute and illustrate the conclusions of Ref.~\cite{jarzynski06-pre}.

In summary, we have offered a detailed path integral study of the Jarzynski equality. This path integral approach allowed us to derive exact results for work-weighted propagators, as well as analytical expressions for typical and dominant trajectories in model problems. As argued above, such propagators are generally not known in closed analytical form, and hence one needs to resort to models such as the present ones for additional insight. Our Euler-Lagrange equations of motion make it possible to investigate quantitative properties of typical and dominant trajectories formerly discussed at a qualitative level. A natural application of such equations can be found in the context of single-molecule ``pulling'' experiments, where it can reveal the typical and dominant rupture events inherent to these processes \cite{hummer05}.

\acknowledgments

The authors are indebted to Attila Szabo for numerous discussions. This research was supported by the Intramural Research Program of the NIH, NIDDK.


\begin{thebibliography}{16}
\expandafter\ifx\csname natexlab\endcsname\relax\def\natexlab#1{#1}\fi
\expandafter\ifx\csname bibnamefont\endcsname\relax
  \def\bibnamefont#1{#1}\fi
\expandafter\ifx\csname bibfnamefont\endcsname\relax
  \def\bibfnamefont#1{#1}\fi
\expandafter\ifx\csname citenamefont\endcsname\relax
  \def\citenamefont#1{#1}\fi
\expandafter\ifx\csname url\endcsname\relax
  \def\url#1{\texttt{#1}}\fi
\expandafter\ifx\csname urlprefix\endcsname\relax\def\urlprefix{URL }\fi
\providecommand{\bibinfo}[2]{#2}
\providecommand{\eprint}[2][]{\url{#2}}

\bibitem[{\citenamefont{Jarzynski}(1997{\natexlab{a}})}]{jarzynski97-prl}
\bibinfo{author}{\bibfnamefont{C.}~\bibnamefont{Jarzynski}},
  \bibinfo{journal}{Phys. Rev. Lett.} \textbf{\bibinfo{volume}{78}},
  \bibinfo{pages}{2690} (\bibinfo{year}{1997}{\natexlab{a}}).

\bibitem[{\citenamefont{Crooks}(1999{\natexlab{a}})}]{crooks99}
\bibinfo{author}{\bibfnamefont{G.~E.} \bibnamefont{Crooks}},
  \bibinfo{journal}{Phys. Rev. E} \textbf{\bibinfo{volume}{60}},
  \bibinfo{pages}{2721} (\bibinfo{year}{1999}{\natexlab{a}}).

\bibitem[{\citenamefont{Crooks}(1999{\natexlab{b}})}]{crooks-thesis}
\bibinfo{author}{\bibfnamefont{G.~E.} \bibnamefont{Crooks}}, Ph.D. thesis,
  \bibinfo{school}{Berkeley} (\bibinfo{year}{1999}{\natexlab{b}}).

\bibitem[{\citenamefont{Chernyak et~al.}(2006)\citenamefont{Chernyak, Chertkov,
  and Jarzynski}}]{chernyak06}
\bibinfo{author}{\bibfnamefont{V.~Y.} \bibnamefont{Chernyak}},
  \bibinfo{author}{\bibfnamefont{M.}~\bibnamefont{Chertkov}}, \bibnamefont{and}
  \bibinfo{author}{\bibfnamefont{C.}~\bibnamefont{Jarzynski}},
  \bibinfo{journal}{J. Stat. Mech.} p. \bibinfo{pages}{P08001}
  (\bibinfo{year}{2006}).

\bibitem[{\citenamefont{Hummer and Szabo}(2001)}]{hummer01-pnas}
\bibinfo{author}{\bibfnamefont{G.}~\bibnamefont{Hummer}} \bibnamefont{and}
  \bibinfo{author}{\bibfnamefont{A.}~\bibnamefont{Szabo}},
  \bibinfo{journal}{Proc. Natl. Acad. Sci. (USA)}
  \textbf{\bibinfo{volume}{98}}, \bibinfo{pages}{3658} (\bibinfo{year}{2001}).

\bibitem[{\citenamefont{Hummer and Szabo}(2005)}]{hummer05}
\bibinfo{author}{\bibfnamefont{G.}~\bibnamefont{Hummer}} \bibnamefont{and}
  \bibinfo{author}{\bibfnamefont{A.}~\bibnamefont{Szabo}},
  \bibinfo{journal}{Acc. Chem. Res.} \textbf{\bibinfo{volume}{38}},
  \bibinfo{pages}{504} (\bibinfo{year}{2005}).

\bibitem[{\citenamefont{Jarzynski}(2006)}]{jarzynski06-pre}
\bibinfo{author}{\bibfnamefont{C.}~\bibnamefont{Jarzynski}},
  \bibinfo{journal}{Phys. Rev. E} \textbf{\bibinfo{volume}{73}},
  \bibinfo{pages}{046105} (\bibinfo{year}{2006}).

\bibitem[{\citenamefont{Zwanzig}(2001)}]{zwanzig01}
\bibinfo{author}{\bibfnamefont{R.}~\bibnamefont{Zwanzig}},
  \emph{\bibinfo{title}{Nonequilibrium Statistical Mechanics}}
  (\bibinfo{publisher}{Oxford Univ. Press}, \bibinfo{address}{New York},
  \bibinfo{year}{2001}).

\bibitem[{\citenamefont{Chaichian and Demichev}(2001)}]{chaichian-book01}
\bibinfo{author}{\bibfnamefont{M.}~\bibnamefont{Chaichian}} \bibnamefont{and}
  \bibinfo{author}{\bibfnamefont{A.}~\bibnamefont{Demichev}},
  \emph{\bibinfo{title}{Path Integrals in Physics, Volume I: Stochastic Process
  and Quantum Mechanics}} (\bibinfo{publisher}{Institute of Physics},
  \bibinfo{address}{London}, \bibinfo{year}{2001}).

\bibitem[{\citenamefont{Adib}(2008)}]{adib08-jpcb}
\bibinfo{author}{\bibfnamefont{A.~B.} \bibnamefont{Adib}}, \bibinfo{journal}{J.
  Phys. Chem. B} \textbf{\bibinfo{volume}{112}}, \bibinfo{pages}{5910}
  (\bibinfo{year}{2008}).

\bibitem[{\citenamefont{Faccioli et~al.}(2006)\citenamefont{Faccioli, Sega,
  Pederiva, and Orland}}]{orland06}
\bibinfo{author}{\bibfnamefont{P.}~\bibnamefont{Faccioli}},
  \bibinfo{author}{\bibfnamefont{M.}~\bibnamefont{Sega}},
  \bibinfo{author}{\bibfnamefont{F.}~\bibnamefont{Pederiva}}, \bibnamefont{and}
  \bibinfo{author}{\bibfnamefont{H.}~\bibnamefont{Orland}},
  \bibinfo{journal}{Phys. Rev. Lett.} \textbf{\bibinfo{volume}{97}},
  \bibinfo{pages}{108101} (\bibinfo{year}{2006}).

\bibitem[{\citenamefont{Sega et~al.}(2007)\citenamefont{Sega, Faccioli,
  Pederiva, Garberoglio, and Orland}}]{orland07}
\bibinfo{author}{\bibfnamefont{M.}~\bibnamefont{Sega}},
  \bibinfo{author}{\bibfnamefont{P.}~\bibnamefont{Faccioli}},
  \bibinfo{author}{\bibfnamefont{F.}~\bibnamefont{Pederiva}},
  \bibinfo{author}{\bibfnamefont{G.}~\bibnamefont{Garberoglio}},
  \bibnamefont{and} \bibinfo{author}{\bibfnamefont{H.}~\bibnamefont{Orland}},
  \bibinfo{journal}{Phys. Rev. Lett.} \textbf{\bibinfo{volume}{99}},
  \bibinfo{pages}{118102} (\bibinfo{year}{2007}).

\bibitem[{\citenamefont{Risken}(1996)}]{risken}
\bibinfo{author}{\bibfnamefont{H.}~\bibnamefont{Risken}},
  \emph{\bibinfo{title}{The Fokker-Planck Equation: Methods of Solution and
  Applications}} (\bibinfo{publisher}{Springer}, \bibinfo{address}{Berlin},
  \bibinfo{year}{1996}), \bibinfo{edition}{2nd} ed.

\bibitem[{\citenamefont{Mazonka and Jarzynski}()}]{mazonka99}
\bibinfo{author}{\bibfnamefont{O.}~\bibnamefont{Mazonka}} \bibnamefont{and}
  \bibinfo{author}{\bibfnamefont{C.}~\bibnamefont{Jarzynski}},
  \eprint{cond-mat/9912121}.

\bibitem[{\citenamefont{Jarzynski}(1997{\natexlab{b}})}]{jarzynski97-pre}
\bibinfo{author}{\bibfnamefont{C.}~\bibnamefont{Jarzynski}},
  \bibinfo{journal}{Phys. Rev. E} \textbf{\bibinfo{volume}{56}},
  \bibinfo{pages}{5018} (\bibinfo{year}{1997}{\natexlab{b}}).

\bibitem[{\citenamefont{Feynman}(1998)}]{feynman98}
\bibinfo{author}{\bibfnamefont{R.~P.} \bibnamefont{Feynman}},
  \emph{\bibinfo{title}{Statistical Mechanics: A Set of Lectures}}
  (\bibinfo{publisher}{Perseus}, \bibinfo{year}{1998}), \bibinfo{edition}{2nd}
  ed.

\end{thebibliography}

\end{document}